# Automating insect monitoring using unsupervised near-infrared sensors

Klas Rydhmer[1,2]*, Emily Bick[1,3], Laurence Still[1], Alfred Strand[1], Rubens Luciano[1], Salena Helmreich[1], Brittany Beck[1], Christoffer Grønne[1], Ludvig Malmros[1], Knud Poulsen[1], Frederik Elbæk[1], Mikkel Brydegaard[1,4,5,6], Jesper Lemmich[1], Thomas Nikolajsen[1]

*[1]FaunaPhotonics APS, Støberigade 14, DK-2450, Copenhagen SV, Denmark.*
*[2]Department of Geosciences and Natural Resource Management, University of Copenhagen, Rolighedsvej 23,1958 Frederiksberg C, Denmark*
*[3] Department of Plant and Environmental Sciences, University of Copenhagen, Frederiksberg C, Denmark*
*[4]Lund laser Centre, Department of Physics, Lund University, Sölvegatan 14, SE-223 62 Lund, Sweden.*
*[5]Center for Animal Movement Research, Department of Biology, Lund University, Sölvegatan 35, SE-223 62 Lund, Sweden.*
*[6]Norsk Elektro Optikk AS, Østensjøveien 34, 0667 Oslo, Norway*
**corresponding author: klry@faunaphotonics.com*

## Abstract

Insect monitoring is critical to improve our understanding and ability to preserve and restore biodiversity, sustainably produce crops, and reduce vectors of human and livestock disease. However, conventional monitoring methods of trapping and identification are time consuming and thus expensive.

Here, we present a network of distributed wireless sensors, recording backscattered near-infrared modulation signatures from insects. The instrument is a compact sensor based on dual-wavelength infrared light emitting diodes and is capable of unsupervised, autonomous long-term insect monitoring over weather and seasons. The sensor records the backscattered light at kHz pace from each insect transiting the measurement volume. Insect observations are automatically extracted and transmitted with environmental metadata over cellular connection to a cloud-based database. The recorded features include wing beat harmonics, melanisation and flight direction.

To validate the sensor's capabilities, we tested the correlation between daily insect counts from an oil seed rape field measured with six yellow water traps and six sensors during a 4-week period. A comparison of the methods found a Spearman's rank correlation coefficient of 0.61 and a p-value=0.0065, with the sensors recording approximately 19 times more insect observations and demonstrating a larger temporal dynamic than conventional trapping.

## Introduction

Insecta is the most speciose class of terrestrial fauna[1] and the majority of the world's biodiversity is composed of this class[2]. In epidemiological and agricultural ecosystems, insects serve as both beneficial organisms[3–5] and economic pests[6,7]. Data on insects can support biodiversity conservation[8,9], human health protection[10] and increased food production[11].

Insects are monitored via established sampling methods including trapping, sweep netting, and portable aspiration [12–14]. These methods are imperfect resulting in biases towards size[15–17] and stage[18]. Additionally, conventional methods may be time-consuming, costly and prone to human error such as person-to-person variation in sampling execution[19–21]. New methods, like insect anesthetization sampling[22], are being implemented to minimize these biases. Regardless of sampling method, insect identification is time consuming and requires specialized training.

In order to reduce the cost of insect monitoring and identification, automation of insect trapping [23–27] and identification [27–31] has been developed. While these methods could greatly improve monitoring via traps, they are unsuitable for monitoring a general insect population since trap designs and baits are generally biased in regard to species.

Automation of insect monitoring without traps could reduce species bias of conventional methods and human error, thus greatly improving the state of the art. Insect identification has been automated as early as 1973 using wingbeat frequency[32–34], and today remote insect sensing includes acoustic detection[35], radar observations[36–38] and lidar[39–41]. Acoustic methods work best with a solid medium[26,42], though acoustic monitoring of free flying insects has been demonstrated[43–45]. While radar technologies have much larger monitoring range[16,38,46–48], they are unsuitable for monitoring small insects, or insects around vegetation, such as a crop canopy. Lidar can be used to record a large number of observations in a long transect[49–53] and distinguish between species groups by wingbeat frequency (WBF)[50,54]. However, lidar equipment requires a trained operator and requires constant supervision due to eye safety restrictions.

Here we present an autonomous near-infrared sensor for monitoring of flying insects in the field. The sensor aims to minimize human biases, be usable by non-technical personnel, and be capable of weatherproof long-term monitoring.

## Instrument design

The sensor is weatherproof, compact, and intended for field use by non-technicians. Like entomological lidar instrumentation, an air volume is illuminated, and light backscattered from insects entering the measurement volume is recorded by a high-speed photodetector. In addition, the instrument is equipped with a satellite navigation device, a camera for situational photos, and an environmental sensor monitoring temperature, humidity, and light intensity. An internal Global System for Mobile Communications (GSM) modem allows for communication and data transfer. The sensor can be powered by any 12V power supply, including utility power, batteries, or solar power, and has a maximum power consumption of 30W during monitoring. A photo of the sensor is shown in Figure 1.

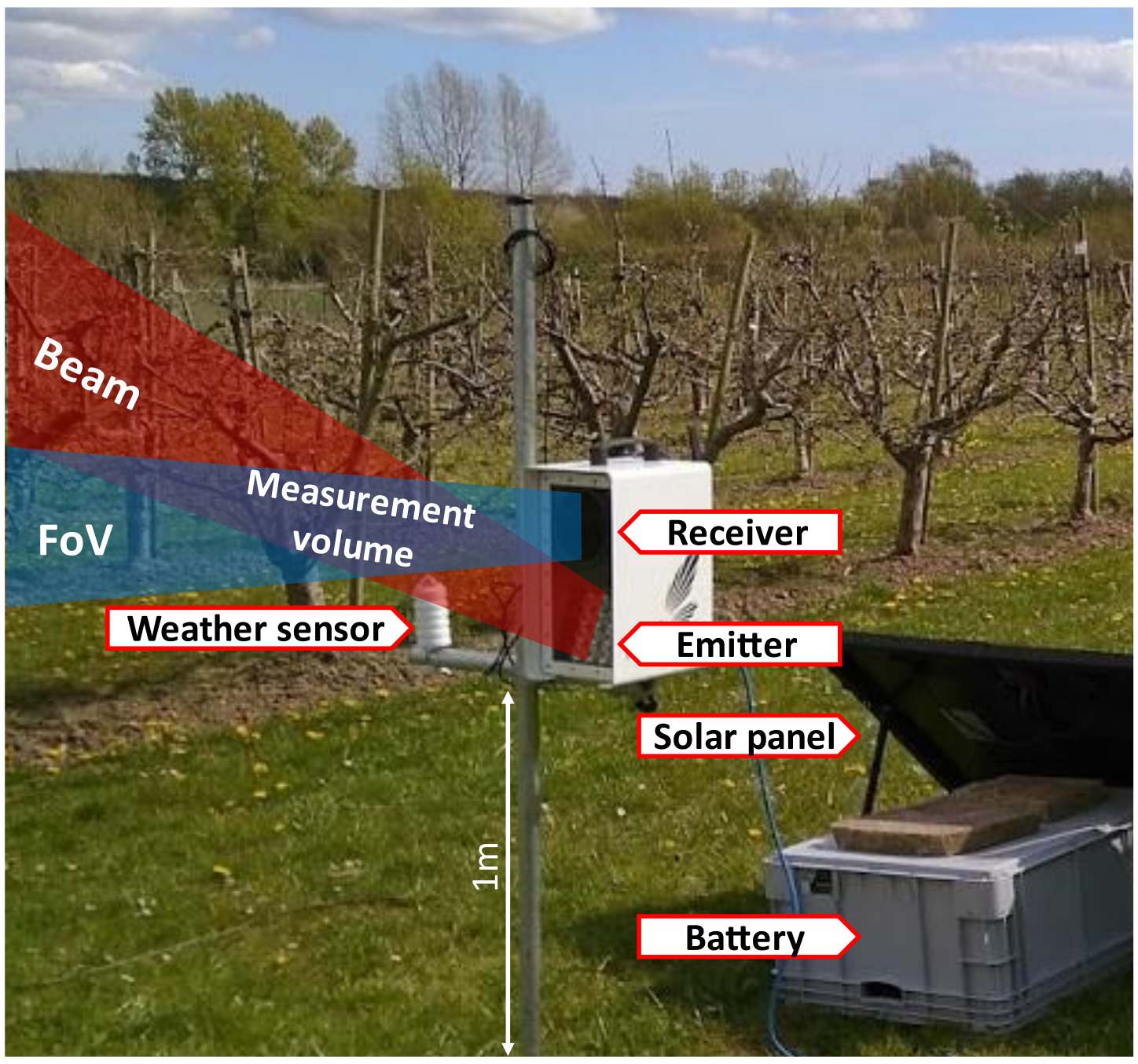

*Figure 1: As insects fly into the measurement volume, the backscattered light is recorded by the receiver. Insect observations are automatically extracted and transmitted along with environmental data, location, and situational photos, to the cloud via a GSM connection. Using a solar panel and battery, the sensor is capable of unsupervised, long-term monitoring in remote locations.*

## Emitter

The emitter module consists of a rectangular array of LEDs emitting two spectral bands at 808 nm and 980 nm with total output of 1.6 W and 1.7 W, respectively. The two wavelengths are modulated in a square wave at 118.8 kHz and 79.2 kHz respectively. The LEDs are mounted in a checkerboard pattern to achieve a homogeneous beam profile. The total area of the checkerboard, and thus the beam size at the source, is 82 cm$^2$. The light emitted from each diode is partially collimated by an asymmetrical lens and expands with 20° and 4° diverging angles ($\theta_E$). The full width half maximum (FWHM) of the emitted light is 26 nm for the 808 nm band and 47 nm for the 970 band.

## Receiver

The backscattered light from insects entering the overlap between the beam and the receiver's field of view (FoV) is collected by a near infrared coated aspheric lens (60 mm focal length, ø 76.2 mm aperture) onto a silicon quadrant photodiode (QPD) with a total area of 1 cm$^2$. The receiver is focused at 1 m and has a 4° divergence angle ($\theta_R$). Quadrant detection of insects allow for basic range and size estimation[55,56] and can differentiate ascending and descending insects as well as migrating insects with tailwind or host- or scent-seeking insects with headwind.

## Signal processing

Each quadrant of the QPD is amplified by a dedicated trans-impedance amplifier (TIA) with a bandwidth of 10 Hz to 1 MHz and a gain of 0.75V/µA around 100kHz. The amplified signals are sampled by four analogue-digital converters (ADC) with 14-bit output at a rate of 6 MHz. The digital data-streams are sent into a field-programmable gate array (FPGA) where eight digital lock-in amplifiers are implemented in VHDL (Very High-Speed Integrated Circuit Hardware Description Language). This allows the two spectral bands to be recorded independently on each quadrant, resulting in an 8-channel data stream. The data is then filtered by a low-pass filter with a cut-off at 5kHz and down-sampled to a 20 kHz, 16-bit data stream before it is sent to a microcontroller unit (MCU) for event extraction and further processing. The increase in bit depth is possible due to the oversampling of the unfiltered signal.

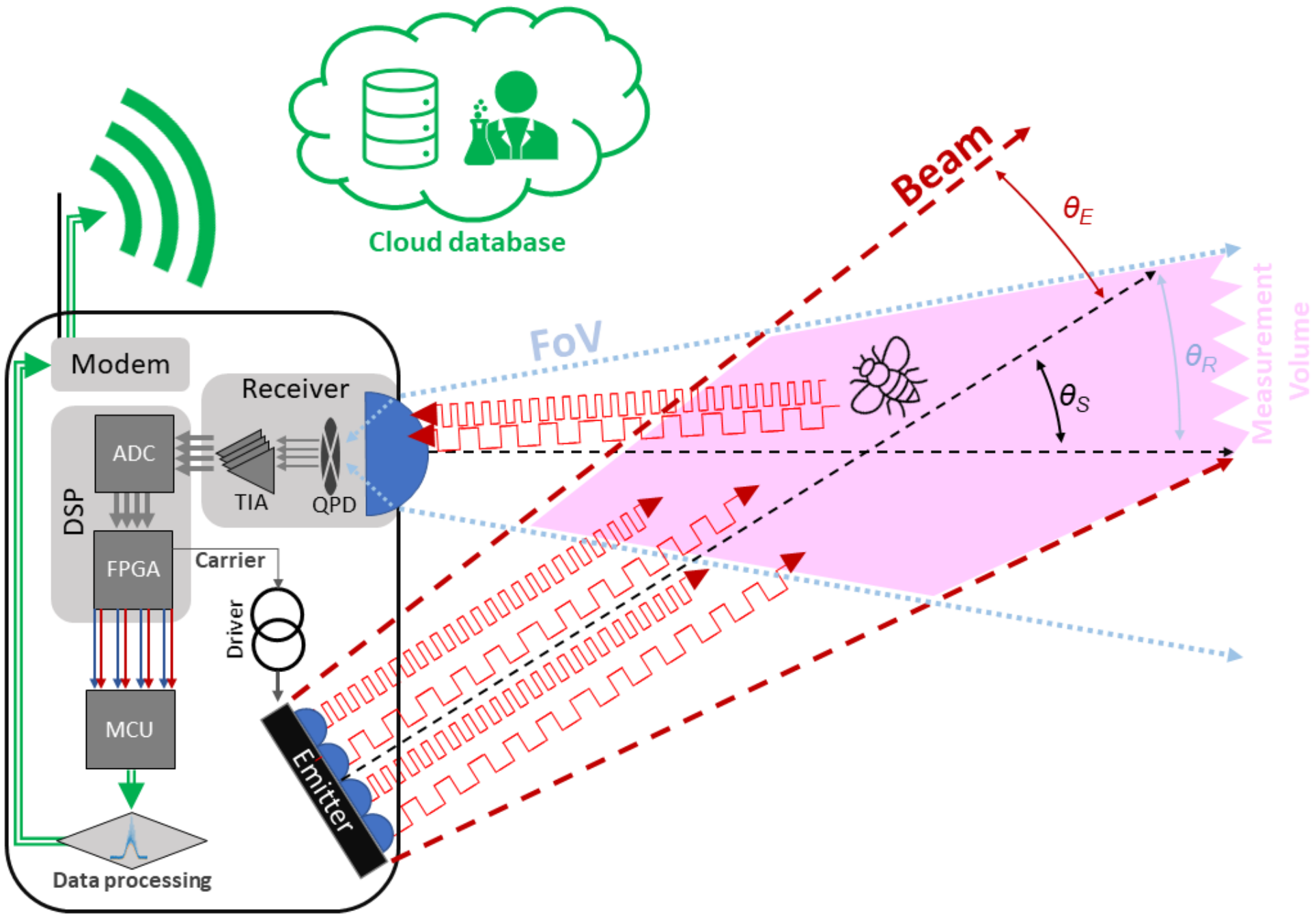

*Figure 2. Light is emitted and collimated from the LED board at 808 nm and 980nm and modulated at different carrier frequencies. The backscattered light from an insect entering the measurement volume is collected by a lens and focused onto a QPD. The four QPD-quadrants are independently amplified by a TIA and sampled. The digital data streams are sent to the FPGA, where 8 digital lock-in amplifiers individually amplify each wavelength in the digital signal processing (DSP) unit. The resulting 8-channel data stream is analyzed by the MCU which extracts events from the data stream. The events can then be stored locally or sent via GSM modem to a cloud database.*

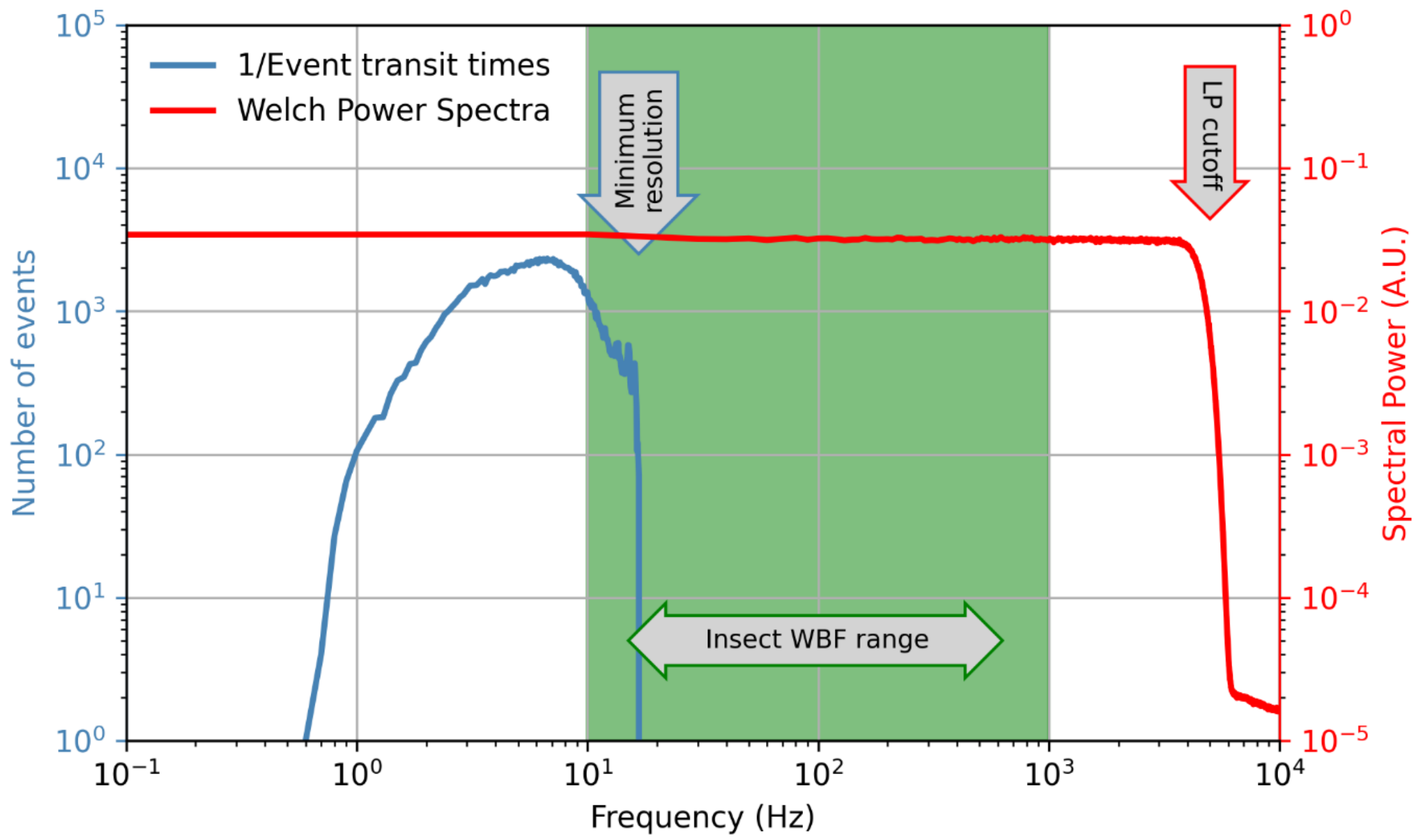

*Figure 3. The wide beam yields long insect transit times, and the corresponding frequency resolution is high enough to accurately capture most species. The frequency response curve (red) is flat in the wingbeat frequency region and the effect of the LP filter at 5 kHz is indicated. The 5kHz bandwidth allows a minimum of 4 harmonic overtones to be recorded even for mosquitoes with very high wingbeat frequencies.*

## Measurement volume

The measurement volume is defined by the overlap between the beam and the FoV. Its size and shape can be adjusted by changing the angle ($\theta_S$) between the emitter and receiver.

The beam, FoV and the measurement volume have been mapped by a custom-built 3-axis robot covering a volume of 2 m x 1.5 m x 1.5 m. The robot is equipped with a photodetector, an illumination source, and a sphere dropping mechanism. By measuring the intensity of the emitted beam and the sensitivity of the FoV in the volume, the signal response from an arbitrary target can be estimated. The volumes were measured at 20 planes along the Z axis, from 30 to 1655 mm, each plane consisting of 56 x 56 measurement points in a 12 mm grid. The calculated signals were then compared to actual measurement values by dropping black and white spheres. The white spheres were assumed to be 100% reflective and the black spheres had a 5% reflectivity.

The measurement volume properties for targets with various optical cross sections (OCS) at different angles are shown in Table 1. The size of the measurement volume is dependent on the minimum acceptable sensitivity, which is related to the noise in the instrument. In the following results, the edge of the volume is defined as the limit where the signal to noise ratio (SNR) is larger than 10 for typical noise levels in a field installation. The volumes for a 10 mm$^2$ target are shown in Figure 4.

| $\theta_S$ (deg) | SNR at 25 cm for 10 mm$^2$ target | Far limit (10 mm$^2$ target) (cm) | Measurement volume for 1 mm$^2$ target (liters) | Measurement volume for 10 mm$^2$ target (liters) | Measurement volume for 100 mm$^2$ target (liters) |
|---|---|---|---|---|---|
| 5 | 725 | > 1650 | 8 | 52 | 100 |
| 12.5 | 1430 | 130 | 7 | 27 | 87 |
| 20 | 1680 | 95 | 5 | 16 | 70 |

*Table 1: Measurement volume parameters at different angles for different target OCS. The target OCS values correspond roughly to a small midge, a small beetle, and a honeybee.*

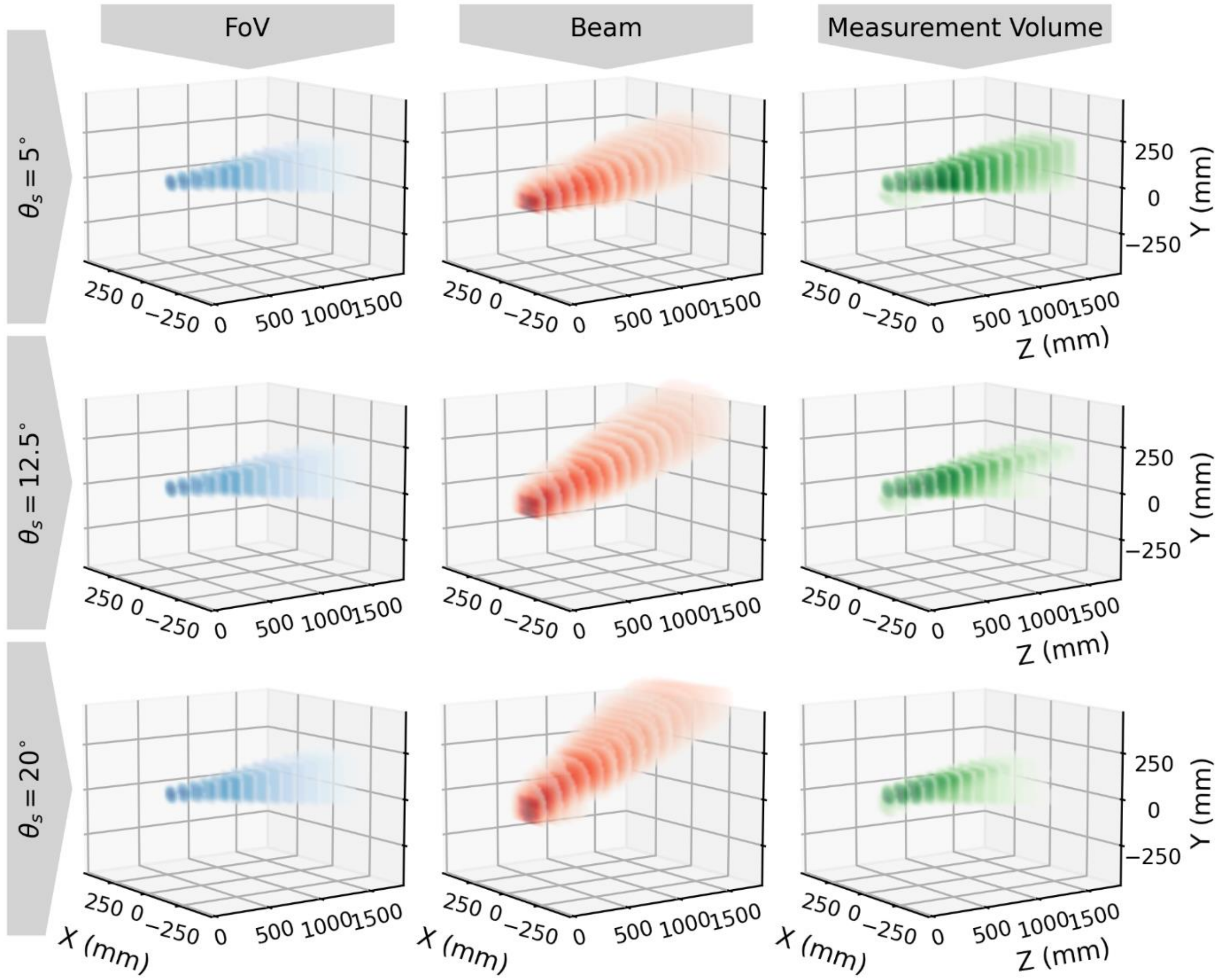

*Figure 4. Measured FoV, beam, and measurement volume for the three angles. Each volume is mapped at 20 planes along the Z axis and each plane consists of 56 x 56 measurement points with 12 mm spacing. For the FoV and beam, all measurement points below 2% of the maximum value are excluded. For the measurement volume all points with a SNR < 10 for a 10 $mm^2$ target are excluded. A low angle yields a longer and larger, but less sensitive, measurement volume. The FoV is identical in all configurations.*

## Data processing

### Automated event extraction

The sensor records intervals of 10 minutes (4 quadrants, 2 spectral bands, 16 bit and 20 kHz sample rate after demux of carrier frequency) and automatically extracts insect observations from each recording. The event extraction is inspired by earlier work but modified to reduce computational load[40,41,54,57]. In each channel, the signal was downsampled to 2 kHz and a rolling median filter with a width of 2 s and 50% overlap was used to estimate the quasi-static baselines (the baselines can change with environmental conditions, static objects in the beam etc.) Similarly, a standard deviation filter with identical properties was applied to all datapoints below the median. This reduces the influence of rare events, such as insects, on the noise level estimation.

The interpolated median signals were removed from the full resolution data and we employed a Boolean condition for insect detection when the time series exceed 10 times the estimated standard deviation. The Boolean time series were eroded by 500 µs and dilated by 30 ms. The erosion rejects short spikes which could not be interpreted and dilation includes insect observation flanks. The logical OR function was applied across all QPD-quadrants and spectral channels. Extracted observations are transmitted to a cloud database along with metadata such as baseline and noise level, via GSM connection or stored locally until a connection is available. An example of the event extraction process is shown in Figure 5, and the insect event is shown in greater detail in Figure 6.

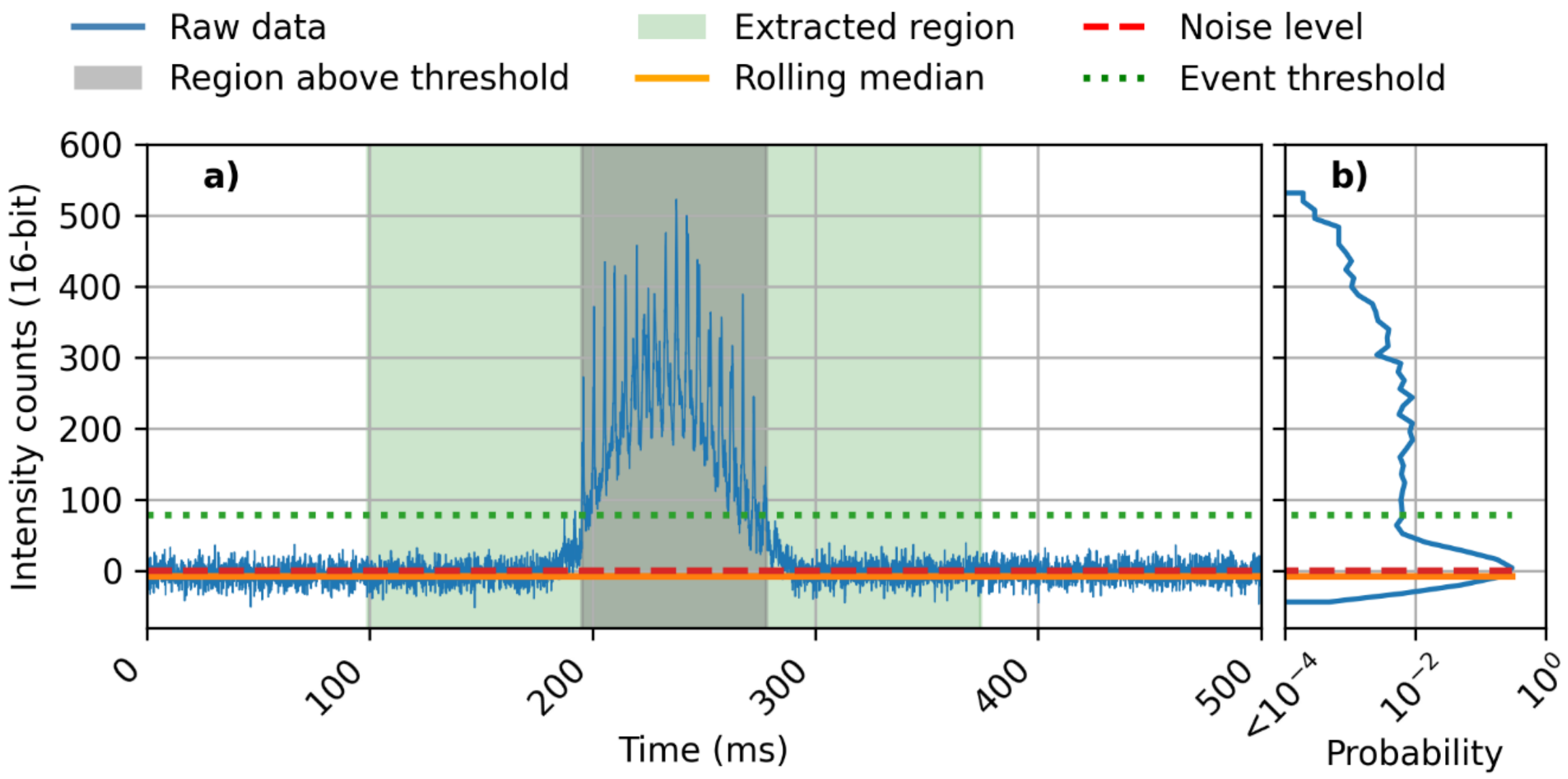

*Figure 5. An example of the event extraction process in a single channel for visibility. a) The data, in the 810 nm band of a single QPD segment after the rolling median has been removed. The part of the signal above the event threshold is marked in grey, and the final insect event after erosion and dilation of the binary map is marked in green. b) Intensity distribution of the data.*

Each insect observation, along with its associated timestamp and device identifier, is automatically uploaded to the cloud via one-way AMQP (Advanced Message Queuing Protocol), with unique connections for each device. Virtual computing is then used to further process, analyze, and securely store data for further use and aggregation.

### Events and Features

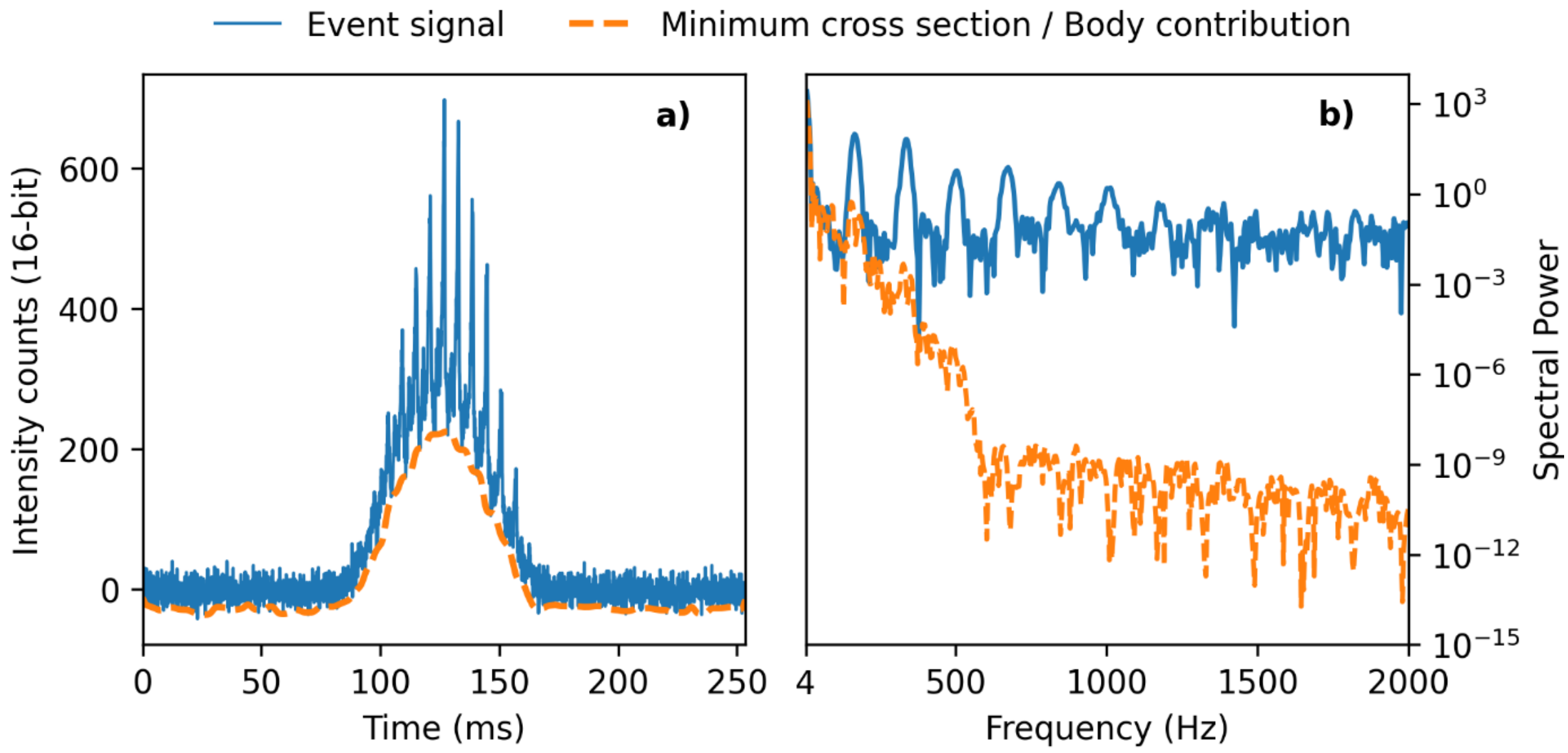

*Figure 6: a) The 810 nm signal for a single insect event in of one of the QPD segments. The insect wingbeats appear as undulating spikes. The minimum envelope of the signal is interpreted as the insect body contribution to the signal. b) The Welch spectral density of the event. The fundamental wingbeat frequency and harmonics are seen in the event signal. This event has a fundamental wingbeat frequency of 160 Hz and an average body-to-wing ratio of 0.4.*

### Feature extraction / Data interpretation

The QPD segments collect backscattered light from different sections of the measurement volume. For a single object passing through the measurement volume, the signal strength within each QPD-quadrant is related to the object's OCS as well as its position. As the OCS varies with each wingbeat, the wingbeat frequency can be resolved. Many methods have been used to extract the wingbeat frequency from insect observations[57–59] and most are based on identifying the fundamental frequency in the frequency domain, as shown in Figure 6 b).

In addition to the wingbeat frequency, the body and wing contribution can be measured from each time signal which allows calculation of additional features such as body-to-wing ratio. Additional features can be calculated by comparing the relative intensity of the body and wing signals in the two spectral bands. These bands differentially index melanin absorption[60–62] and may yield some sensitivity to wing interference patterns[61,63,64], although not enough to uniquely determine wing membrane thickness. Together these features can be used to quantify the morphology of different insect groups and allow remote classification of insects according to order, family, genus or species[59,64–66].

## Field validation

### Methodology

The sensor was field-tested against a conventional insect monitoring method, yellow water traps (22 cm diameter)[67,68], in an organic oilseed rape (*Brassica napus* L.) field in Sorø, Denmark (55°29'04.3"N 11°29'34.6"E). During a four-week period (04/22/19 - 05/22/19), insects were monitored with six sensors and six yellow water traps. Sensors and traps were placed in a grid pattern, consisting of four linear transects 30 m from and perpendicular to the field's southern-most edge. Each transect consisted of three monitoring points (either sensors or traps) with 45 m spacing, and a separation of 22.5 m between transects. The first and third transect consisted of sensors and the second and fourth were yellow water traps. During the field study presented in this work, $\theta_S$ was set to 20° in order to maximize the signal strength of small targets at close range.

Fundamentally the two methods observe different insect behaviors. While the sensor looks at insects flying above the crop canopy, the yellow water traps look at insects that occur within it. Further confounding the comparison, yellow is attractive to some insects[68]. Therefore, some proportion of insects will be attracted to the yellow water traps, resulting in overrepresentation of some species[69,70].

## Data analysis

The water traps were emptied daily, and sensor data was recorded continuously. All insects in the traps were collected, but to allow for a more direct comparison of methods, non-flying insects and thrips found in water traps were excluded from further analysis.

The sensor data was aggregated according to the collection time of the water traps but one day, April 30th, was excluded due to instrument malfunction. The average number of recorded insect observations per sensor per day and per hour was calculated. The calculated numbers were normalized by sensor uptime, which was on average 90% throughout the measurement period.

## Results

The insect activity recorded by the sensors and traps respectively are shown in Figure 7. Insect counts from sensors and traps cannot be directly equated due to differences in measurement subject (insect flights vs insect landings) and non-homogeneous insect distribution; however, they serve to visualize similarities in gross changes in insect activity over the sample period. The results demonstrate a significant correlation between the sensor and trap results, specifically with a Spearman's rank correlation coefficient of 0.61 and a p-value=0.0065 [71]. Over the course of the season, an average of 1122 ± 242 (SE) insect observations per day were collected per sensor (excluding downtime), compared to an average of 63 ± 6 (SE) insects caught per water trap per day over the same period.

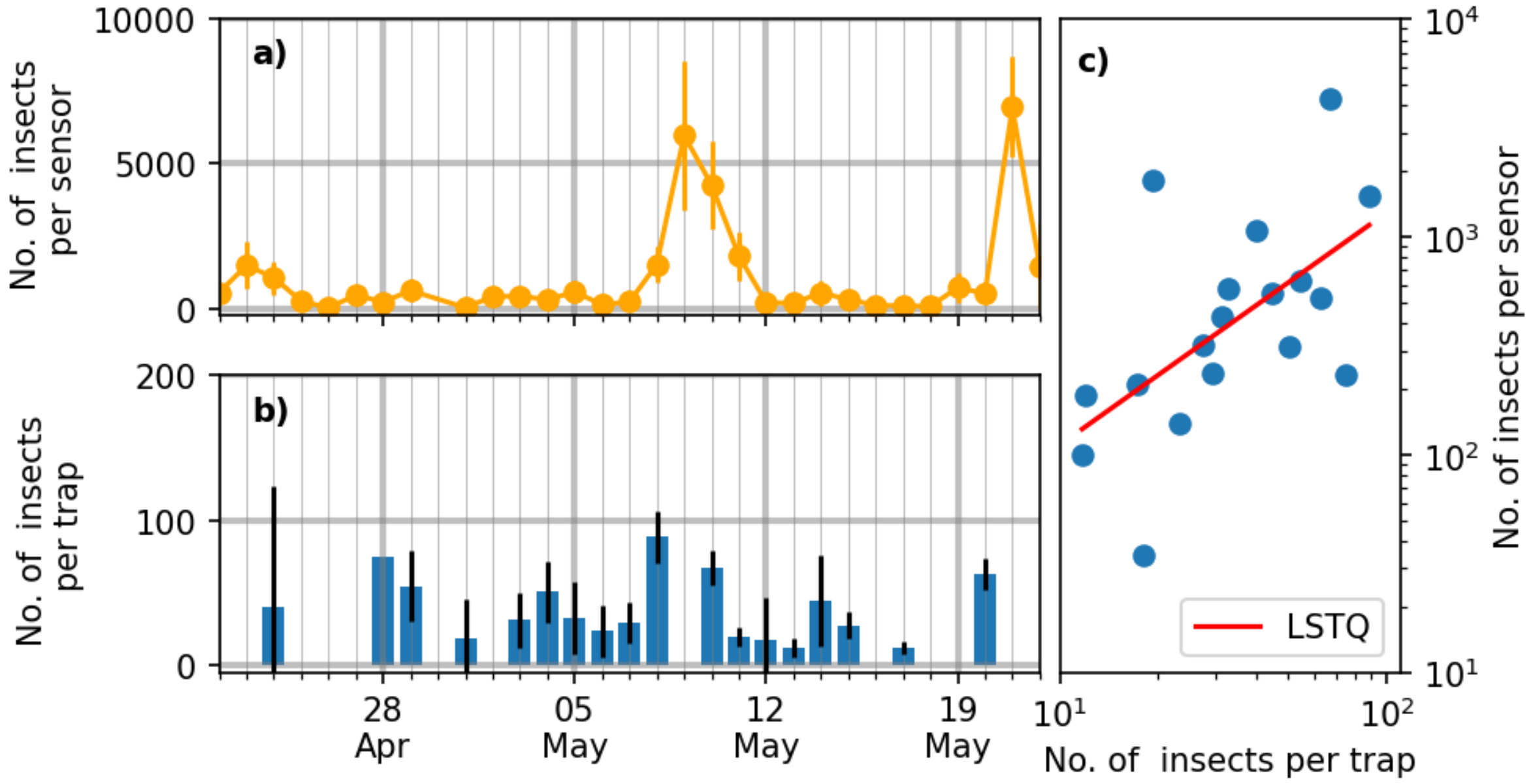

*Figure 7. a) Average insect counts across sensors per day b) Average insect count across yellow traps per day. c) Sensor vs trap counts during days where both sensor and trap data was available. The red line is the linear least square fit (LSTQ) with a Spearman correlation coefficient of 0.61.*

## Discussion

Here we present a sensor for automated unsupervised field monitoring of insect flight activity. The sensor illuminates an air volume and records the backscattered light from insects that fly through the measurement volume. Discrete insect observations are automatically extracted from the continuous raw data flow and transmitted over a cellular connection to a database in the cloud. Field validation showed the number of recorded insect observations correlates with the number of individual insects trapped by a conventional insect monitoring method. Furthermore, the sensor recorded an order of magnitude more insects than the conventional method over the same time period.

The automation of insect monitoring has the potential to reduce monitoring bias, cost, and human labor, potentially resulting in an increased ability to collect large quantities of biodiversity, public health, and economically relevant insect data. Additionally, the observations from the sensors were available in real time, whereas emptying and counting insects from traps required a significant amount of labor. While this work was limited to comparing total insect counts from the traps, it is possible for a skilled expert to identify these insects to the sub-species level. This is an area were the traps have a strong advantage over this sensor and similar instrumentation. However, we expect significant progress with species identification from sensor observations.

One of the most striking differences in monitoring methods is the day-to-day variability in the number of data points collected (Figure 7). While the yellow traps catch a similar number of insects each day, the difference between low and high flight activity days were more visible in the sensor. Early analysis of the trap and sensor data indicates that the peak recorded during May 7-11 is due to a pollen beetle (*Brassicogethes aeneus*) activity spike. This will be the subject of further studies.

Another marked difference between the sensor and the water traps is the number of data points collected over the same collection period. Each sensor observed ~ 19 x more insect observations than insects collected in the water trap. While in general the correlation between the two values is considered more relevant than the absolute number, one advantage of a much higher observation rate in the sensors is an increased probability of recording rare species. Further work is required to successfully identify rare species from sensor signals, but if realized, sensors could provide a method to monitor rare species across large areas with lower labor costs.

Insect observations recorded by the sensor are precisely timestamped at the point of occurrence, allowing variable aggregation over time as well as higher time granularity in measurements. This may have advantages in standardization compared to conventional trapping periods which may be affected by variable intra-collection times due to human factors (e.g. missed or delayed collections due to weekends or public holidays). Furthermore, the higher granularity and continuous monitoring during unsociable hours allows for the comparatively easy and low-labor collection of data on insect circadian rhythms, as well as direct weather interactions.

We hypothesize that the sensors observe different insect behaviors compared to conventional monitoring methods since only airborne (flying or jumping) insects are recorded. Therefore, we did not expect a perfect correlation between the sensors and the conventional methods. Sweep netting is likely the most similar monitoring method since it also catches insects in flight above the crop. However, sweep netting, which also collects insects on plants, occurs at a point measurement in time and is typically performed along a transect, rather than at a fixed point in the field[19]. Also, each trapping method is biased towards different insects, influencing catch[15,17].

Trapping methods, such as the water traps used in this study, monitor insects landing, walking, or jumping to a specific point and do not record insects in flight. Also, each trapping method is biased towards different insects, with the trap color influencing the trap catch[68]. Although we do not yet know in what manner, the sensor is also most likely biased towards certain species groups. Most primarily, its only capable of recording airborne insects. Insect vision is focused towards the visual or ultraviolet spectrum and not capable of resolving infrared light and we believe the emitted beam has very little influence on insect behavior[72]. However, in a homogeneous landscape such as an agricultural field, any foreign object could serve as an attractant and the placement above the canopy could attract insects. Finally, the size of the measurement volume varies with the OCS of the insects and larger insects will be over-represented. To provide a complete picture of the insect population, this should be considered. Along with species specific observations, this is an area where we expect significant progress.

The automated unsupervised field sensor has the potential to facilitate pest prevention, public health studies and biodiversity monitoring. In further work we will explore the possibilities of unsupervised long-term monitoring of insect activity and species recognition.

## Conclusions

In this work, we have introduced an unsupervised automated sensor for insect monitoring. The measurement principle is similar to entomological lidar setups but is optimized for near-field measurements. This simplifies the installation process and increases the robustness of the sensor, allowing it to be operable by non-technical experts and enables long-term unsupervised monitoring.

The sensor automatically extracts insect events from the raw data and transmits these via a built-in modem for further processing. From the recorded observations, features such as the wingbeat frequency, body-wing ratio, and melanisation factor are computed and used to predict the insect classification down to species. During a 4-week deployment in an oilseed rape field, the detected flight activity was shown to be correlated with a conventional monitoring method.

The capabilities and scalability of this sensor-based method has the potential to improve the state of the art in insect monitoring. The sensor can be used to explore areas such as biodiversity assessment, insecticide resistance, and long-term monitoring of remote areas, facilitating research studies currently difficult or impossible to conduct with conventional methods.

## Acknowledgments

The authors want to thank Jakob Dyhr for kindly making his organic oilseed rape field in Sorø, Denmark, available for the field experiments. Thanks to Lene Sigsgaard, Samuel Jansson and Sam Cook for helpful discussions.

## Funding

This work was supported by Innovation Fund Denmark under grant no. 9078-00183B and the Danish Environmental Protection Agency under grant no. MST-667-00253.

## Author contributions:

KR Wrote the first draft, produced figures, and conducted data analysis. KR, EB, and LS developed paper outline and structure. EB contributed to the introduction, field validation, discussion, and conclusion. LS contributed to the data processing section and discussion. KP and LM contributed to the instrument software development. AS, RL, and FE contributed to the instrument development and instrument characterization section.

MS contributed with editing and contributed to figures.

SH, BB, CG & JL collected and counted insects during the field trials.

TN led the development of the instrumentation. JL took over development leadership in 2020.

## Competing Interests

All authors are or were (partly) affiliated with FaunaPhotonics, the company that developed the sensor described in this study.